\documentclass[]{article}

\usepackage[nottoc,notlot,notlof]{tocbibind}
\usepackage{graphicx}
\usepackage{tabularx}
\usepackage{multirow, multicol}
\usepackage{subcaption}
\usepackage{afterpage}
\usepackage{amsmath,amssymb,amsthm}
\usepackage{rotating}
\usepackage{fancyhdr}
\usepackage{tikz}
\usepackage{rotating}
\usepackage{float}
\usepackage[pdfusetitle, colorlinks, linkcolor=black, citecolor=black, urlcolor=black]{hyperref}
\usepackage{algorithm}
\usepackage{algpseudocode}

\usepackage[english]{babel}

\newtheorem{theorem}{Theorem}

\newtheorem{lemma}{Lemma}
\newtheorem{definition}{Definition}

\newtheorem{proposition}{Proposition}
\newtheorem{observation}{Observation}

\title{Machine Learning Techniques for Stackelberg Security Games: a Survey}
\author{Giuseppe De Nittis, Francesco Trov\`o \\ \\ Politecnico di Milano \\ Dipartimento di Elettronica, Informazione e Bioingegneria \\ \{giuseppe.denittis, francesco1.trovo \}@polimi.it }
\date{\today}
\begin{document}
\maketitle
\ \\
\ \\
\ \\
\begin{abstract}
The present survey aims at presenting the current machine learning techniques employed in security games domains. Specifically, we focused on papers and works developed by the Teamcore of University of Southern California, which deepened different directions in this field. After a brief introduction on Stackelberg Security Games (SSGs) and the poaching setting, the rest of the work presents how to model a boundedly rational attacker taking into account her human behavior, then describes how to face the problem of having attacker's payoffs not defined and how to estimate them and, finally, presents how online learning techniques have been exploited to learn a model of the attacker.
\end{abstract}

\clearpage
\tableofcontents

\section{Introduction}
The present survey aims at presenting the current main machine learning techniques employed in security games domains. Specifically, we focused on papers and works developed by the Teamcore of University of Southern California, which deepened different directions in this field. Among several works on this topic, e.g.,~\cite{fang2015security, nguyen2016capture, sinha2015learning, qianonline, nguyenaddressing, nguyen2016conquering}, this paper is essentially based on these works~\cite{nguyen2013analyzing, kar2015game, haghtalabthree, nguyen2015making, qian2016restless, qian2014online}.

After a brief introduction on Stackelberg Security Games (SSGs) and the poaching setting, the rest of the work is organized according to the different problems that has been dealt with.

\begin{itemize}
\item Section~\ref{sec:suqr} shows how to model a boundedly rational attacker taking into account her human behavior.
\item Section~\ref{sec:mmr} faces the problem of having attacker's payoffs not defined and how to estimate them studying the regret of the defender.
\item Section~\ref{sec:mab} presents how online learning techniques have been exploited to learn a model of the attacker.
\end{itemize}

\subsection{Stackelberg paradigm and SSG}
Usually, to represent security scenarios, a specific class of games is adopted, i.e., Stackelberg Games~\cite{stackelberg1934duopoly}. Here, on one side there is the Defender, which publicly commits to a mixed strategy, i.e., a probability distribution on the actions available to the player, and on the other side there is an Attacker, that observes the commitment of the Defender and acts consequently. Such games are called Stackelberg Security Games (SSG). Specifically, in SSGs, the Defender attempts to protect a set of $T$ targets from an Attacker, by optimally allocating a set of $R$ resources, $R < T$. Denote by $\boldsymbol{x} = \{x_t\}$ the Defender's strategy where $x_t$ is the coverage probability at target $t$, the set of feasible strategies is: $\boldsymbol{X} = \{\boldsymbol{x} : 0 \leq x_t \leq 1, \sum_t x_t \leq R\}$. If the adversary attacks $t$ when the defender is not protecting it, the adversary receives a reward $R^a_t$, otherwise the adversary gets a penalty $P^a_t$. Conversely, the Defender receives a penalty $P^d_t$ in the former case and a reward $R^d_t$ in the latter case. Let $(\boldsymbol{R}^a, \boldsymbol{P}^a)$ and $(\boldsymbol{R}^d, \boldsymbol{P}^d)$ be the payoff vectors. The players' expected utilities at $t$ is computed as:

\begin{align*}
	& U^a_t(x, R^a, P^a) = x_t P^a_t + (1-x_t) R^a_t\\
	& U^d_t(x, R^d, P^d) = x_t R^d_t + (1-x_t) P^d_t
\end{align*}

In general, in such situations, the most appropriate solution concept is the Leader--follower equilibrium. The problem of finding such equilibrium can be formulated as:
\begin{equation*}
\begin{array}{rllll}
\arg \max\limits_{x_{\textsf{l}},x_{\textsf{f}}^*}	&	\sum\limits_{a_{\textsf{l}} \in A_{\textsf{l}}} \sum\limits_{a_{\textsf{f}} \in A_{\textsf{f}}} \left[U_{\textsf{l}} (a_{\textsf{l}}, a_{\textsf{f}})  x_{\textsf{l}}(a_{\textsf{l}})  x_{\textsf{f}}^*(a_{\textsf{f}})\right]				\\
\text{s.t.}					&	\sum\limits_{a_{\textsf{l}} \in A_{\textsf{l}}}x_{\textsf{l}}(a_{\textsf{l}})	= 1 &																																		\\
						&	x_{\textsf{l}}(a_{\textsf{l}}) \geq 0			&			&	\forall a_{\textsf{l}}\in A_{\textsf{l}}																		\\
						x_f^*\in\arg\max\limits_{x_{\textsf{l}}} 			&	\sum\limits_{a_{\textsf{l}} \in A_{\textsf{l}}}\sum\limits_{a_{\textsf{l}} \in A_{\textsf{l}}} \left[U_{\textsf{f}} (a_{\textsf{f}}, a_{\textsf{l}})  x_{\textsf{f}}(a_{\textsf{f}}) x_{\textsf{l}}(a_{\textsf{l}})\right]	& &					\\
						\text{ s.t.}	&	\sum\limits_{a_{\textsf{f}} \in A_{\textsf{f}}}x_{\textsf{f}}(a_{\textsf{f}})	= 1					&		&					\\
						& x_{\textsf{f}}(a_{\textsf{f}}) \geq 0 & & \forall a_{\textsf{f}}\in A_{\textsf{f}}							\\
\end{array}
\end{equation*}

where $x_l(a_l)$ ($x_f(a_f)$) is the probability that the leader (follower) will play action $a_l$ ($a_f$) and $x_f^*$ is the best strategy of the follower.

In zero--sum games, the Leader--follower equilibrium coincides with the Nash equilibrium~\cite{simaan1973stackelberg} and the maxmin/minmax strategies.

\subsection{The poaching setting}
Poaching and illegal over--fishing are critical international problems leading to destruction of ecosystems. For example, three out of nine tiger species have gone extinct in the past $100$ years and others are now endangered due to poaching~\cite{secretariat2013global}. Law enforcement agencies in many countries are hence challenged with applying their limited resources to protecting endangered animals and fish stocks.

Building upon the success of applying SSGs to protect infrastructure
including airports~\cite{pita2008deployed}, ports~\cite{shieh2012protect} and trains~\cite{yin2012trusts}, researchers are now applying game theory to green security domains, e.g., protecting fisheries from over--fishing~\cite{brown2014addressing, haskell2014robust} and protecting wildlife from poaching~\cite{yang2014adaptive}. There are several key features in green
security domains that introduce novel research challenges.

\begin{enumerate}
\item The defender is faced with multiple adversaries who carry out repeated and frequent illegal activities (attacks), yielding a need to go beyond the one--shot SSG model.
\item In carrying out such frequent attacks, the attackers generally do not conduct extensive surveillance before performing an attack and spend less time and effort in each attack, and thus it becomes more important to model the attackers' bounded rationality and bounded surveillance.
\item There is more attack data available in green security domains than in infrastructure security domains, which makes it possible to learn the attackers' decision making model from data.
\end{enumerate}

\section{Using human behavior models in solving SSGs}\label{sec:suqr}
In game theory, the adversary is usually represented as a fully rational player. In real--world, people are not fully rational, i.e., their choices are not simply determined by mere calculations. The problem that is addressed here is the representation of a boundedly rational attacker in SSGs. In fact, the fully rationality condition is relaxed to make an important step towards modeling real--world attacker, both to extract useful information about already gathered data, but also to build new models of attackers to enhance the current level of security.

\subsection{SUQR: modeling  a boundedly rational attacker}
In SSGs, attacker bounded rationality is often modeled via behavior models such as Quantal Response (QR)~\cite{mckelvey1995quantal}. The QR model predicts a stochastic distribution of the adversary response: the greater the expected value of a target the more likely the adversary will attack that target. QR's key parameter $\lambda$ represents the level of rationality in adversary's response: as $\lambda$ increases, the predicted response by the QR model converges to the optimal action of the adversary. Instead of using a human behavior model, MATCH, the best algorithm up to 2013, computes a robust defender strategy by guaranteeing a bound on the defender's loss in her expected value if the adversary deviates from her optimal choice. More specifically, the defender's loss is constrained to be no more than a factor of $\beta$ times the adversary's loss in his expected value. The key parameter $\beta$ describes how much the defender is willing to sacrifice when the adversary deviates from the optimal action.

The key idea in Subjective Expected Utility (SEU) is that individuals have their own evaluations of each alternative during decision making.
Recall that in an SSG, the information presented to the human subject for each choice includes: the marginal coverage on target $t (x_t)$; the subject's reward and penalty $(R^a_t, P^a_t)$; the defender's reward and penalty $(R^d_t, P^d_t)$. Inspired by the idea of SEU, it has been proposed~\cite{nguyen2013analyzing} a subjective utility function of the adversary for SSG as the following:

\begin{equation*}
\hat{U}^a_{t'}(\boldsymbol{x}, \boldsymbol{R^a}, \boldsymbol{P^a}) = w_1x_t + w_2R^a_t + w_3P^a_t
\end{equation*}

In fact, SUQR is motivated by the lens model which suggested that evaluation of adversaries over targets is based on a linear combination of multiple observable features. One key advantage of these behavioral models is that they can be used to predict attack frequency for multiple attacks by the adversary, where in the attacking probability is a normalization of attack frequency.

The novelty of this subjective utility function is the linear combination of the values (rewards/penalty) and coverage probabilities. (Note that the decision--making of the general population is modeled, not of each individual since there are no sufficient data for each specific subject). This model actually leads to higher prediction accuracy than the classic expected value function. Other alternatives to this subjective utility function are feasible, e.g., including all the information presented to the subjects:

\begin{equation*}
\hat{U}^a_{t}(\boldsymbol{x}, \boldsymbol{R^a}, \boldsymbol{P^a}) = w_1x_t + w_2R^a_t + w_3P^a_t + w_4R^d_t + w_5P^d_t
\end{equation*}

Then, the QR model is modified by replacing the classic expected value function with the SU function, leading to the SUQR model. In the SUQR model, the probability that the adversary chooses target $t$, $q_t$, is given by:

\begin{equation*}
q_t(\boldsymbol{x}, \boldsymbol{R^a}, \boldsymbol{P^a}) = \frac{e^{\lambda \hat{U}^a_t(\boldsymbol{x}, \boldsymbol{R^a}, \boldsymbol{P^a})}}{\sum_{t'}e^{\lambda \hat{U}^a_{t'}(\boldsymbol{x}, \boldsymbol{R^a}, \boldsymbol{P^a})}}
\end{equation*}

The problem of finding the optimal strategy for the defender can therefore be formulated as:

\begin{align*}
& max_x \sum_{t=1}^T \frac{e^{\lambda U^a_t(\boldsymbol{x}, \boldsymbol{R^a}, \boldsymbol{P^a})}}{\sum_{t'}e^{\lambda U^a_{t'}(\boldsymbol{x}, \boldsymbol{R^a}, \boldsymbol{P^a})}} (x_tR^d_t + (1-x_t)P^d_t)\\
& \text{s.t.} \sum_{t=1}^T x_t \leq K, 0 \leq x_t \leq 1 
\end{align*}

Here, the objective is to maximize the defender's expected value given that the adversary chooses to attack each target with a probability according to the SUQR model.

\subsubsection{Learning SUQR parameters}
Without loss of generality, $\lambda$ = 1. As customarily done in traditional machine learning, Maximum Likelihood Estimation (MLE) to learn the parameters $(w_1, w_2, w_3)$ is employed. Given the defender strategy $\boldsymbol{x}$ and $N$ samples of the players' choices, the log--likelihood of $(w_1, w_2, w_3)$ is given by:

\begin{equation*}
logL(w_1, w_2, w_3| \boldsymbol{x}) = \sum_{j=1}^N log[q_{t_j}(w_1, w_2, w_3| \boldsymbol{x})]
\end{equation*}

where $t_j$ is the target that is chosen in sample $j$ and $q_{t_j}(w_1, w_2, w_3| \boldsymbol{x})$ is the probability that the adversary chooses the target $t_j$. Let $N_t$ be the number of subjects attacking target $t$. Then:

\begin{equation*}
logL(w_1, w_2, w_3| \boldsymbol{x}) = \sum_{t=1}^T N_tlog[q_{t_j}(w_1, w_2, w_3| \boldsymbol{x})]
\end{equation*}

It can be shown that the Hessian matrix of $logL[q_{t_j}(w_1, w_2, w_3| \boldsymbol{x})]$ is negative semi--definite. Thus, this function has an unique local maximum point and hence a convex optimization solver can be used to compute the optimal weights $(w_1, w_2, w_3)$, e.g., fmincon in \verb+Matlab+.

\subsection{From MATCH to SHARP}
Starting from the above results, a new model is now introduced, SHARP~\cite{kar2015game}, which:

\begin{itemize}
\item reasons based on success or failure of the adversary's
past actions on exposed portions of the attack surface to model adversary adaptiveness;
\item reasons about similarity between exposed and unexposed areas of the attack surface, and also incorporates a discounting parameter to mitigate adversary's lack of exposure to enough of the attack surface;
\item  integrates a non--linear probability weighting function to capture the adversary's true weighting of probability.
\end{itemize}

Following the approach presented in the previous section, MLE has been applied to learn the weights of the SUQR model based on data collected from our human subject experiments and found that the weights on coverage probability were positive for all the experiments. That is, counter--intuitively humans were modeled as being attracted to regions with high coverage probability, even though they were not attacking targets with very high coverage but they were going after targets with moderate to very low coverage probability.

To address this issue, a solution can be the augmentation of the Subjective Utility function with a two--parameter probability weighting function, that can be either inverse S--shaped (concave near probability zero and convex near probability one) or S--shaped.

\begin{equation}
f(p) = \frac{\delta p^{\gamma}}{\delta p^{\gamma} + (1-p)^{\gamma}}
\end{equation}\label{eq:fp}

The SU of an adversary denoted by $a$ can then be computed as:

\begin{equation*}
SU_i^a(x) = w_1f(x_i) + w_2R_i^a + w_3P^a_i,
\end{equation*}

where $f(x_i)$ for coverage probability $x_i$ is computed as per Equation (1). The two parameters $\delta$ and $\gamma$ control the elevation and curvature of the function respectively. $\gamma < 1$ results in an inverse S-shaped curve while $\gamma > 1$ results in an S--shaped curve. This is the PSU (Probability weighted Subjective Utility) function. The curve representing human weights for probability is S--shaped in nature, and not inverse S-shaped as prospect theory suggests. The S--shaped curve indicates that people would overweigh high probabilities and underweigh low to medium probabilities.

W.r.t. MATCH, SHARP introduces a new feature –-- distance --– that affects the reward and hence the obvious question for us was to investigate the effect of this new feature in predicting adversary behavior. Several variations of PSU with different combinations of features can be considered.

\begin{align} 
&SU_i^a(x) = w_1f(x_i) + w_2\phi_i + w_3P^a_i\\
&SU_i^a(x) = w_1f(x_i) + w_2R^a_i + w_3P^a_i + w_4D_i\\
&SU_i^a(x) = w_1f(x_i) + w_2\phi_i + w_3P^a_i + w_4D_i
\end{align}\label{eq:psu_examples}

where $\phi_i$ and $D_i$ refer to the animal density at target $i$ and the distance to target $i$ from the poacher's starting location respectively.

\subsubsection{Adaptive utility model}
A second major innovation in SHARP is the adaptive nature of the adversary and addressing the issue of attack surface exposure. The attack surface $\alpha$ is defined as the n--dimensional space of the features used to model adversary behavior.

For example, as per the third PSU model in Equation (4), this would mean the space represented by the following four features: coverage probability, animal density, adversary penalty and distance from the starting location.

A target profile $\beta_k \in \alpha$ is defined as a point on the
attack surface $\alpha$ and can be associated with a target. Exposing the adversary to a lot of different target profiles would therefore mean exposing the adversary to more of the attack surface and gathering valuable information about their behavior. While a particular target location, defined as a distinct region in the 2--d space, can only be associated with one target profile in a particular round, more than one target may be associated with the same target profile in the same round.

\begin{observation}
Adversaries who have succeeded in attacking a target associated with a particular target profile in one round, tend to attack a target with similar target profiles in next round.
\end{observation}

\begin{observation}
Adversaries who have failed in attacking a target associated with a particular target profile in one round, tend not to attack a target with ‘similar' target profiles in the next round.
\end{observation}

The vulnerability associated with a target profile $\beta_i$, which was shown to the adversary in round $r$, denoted $V^r_{\beta_i}$, is defined as a function of the total number of successes and failures on the concerned target profile in that round (denoted by $success^r_{\beta_i}$ and $failure^r_{\beta_i}$ respectively).

\begin{equation*}
V^r_{\beta_i} = \frac{success^r_{\beta_i}-failure^r_{\beta_i}}{success^r_{\beta_i}+failure^r_{\beta_i}}
\end{equation*}

Therefore, more successful attacks and few failures on a target profile indicate that it was highly vulnerable in that round. Because multiple targets can be associated with the same target profile and the pure strategy generated based on the mixed strategy $x$ in a particular round may result in a defender being present at some of these targets while not at others, there may be both successes and failures associated with the same target profile in that round.

The attractiveness of a target profile $\beta_i$ at the end of round $R$, denoted $A^R_{\beta_i}$, is defined as a function of the vulnerabilities for $\beta_i$ from round 1 to round $R$.

\begin{equation*}
A^R_{\beta_i} = \frac{\sum_{r=1}^R V^r_{\beta_i}}{R}
\end{equation*}

Therefore, the attractiveness of a target profile is modeled as the average of the vulnerabilities for that target profile over all the
rounds till round $R$. This is consistent with the notion that a target profile which has led to more successful attacks over several rounds will be perceived as more attractive by the adversary.

\subsubsection{SHARP's utility Computation} 
Existing models (such as SUQR) only consider the adversary's actions from round $(r - 1)$ to predict their actions in round $r$. However, it is clear that the adversary's actions in a particular round are dependent on his past successes and failures. Thus, a novel adaptive probability weighted subjective utility function that captures this adaptive nature of the adversary's behavior by capturing the shifting trends in attractiveness of various target profiles over rounds is proposed.

\begin{align*}
ASU^R_{\beta_i} &= (1-d*A^R_{\beta_i})w_1f(x_{\beta_i}) + (1+d*A^R_{\beta_i})w_2\phi_{\beta_i} +\\
&+(1+d*A^R_{\beta_i})w_3P^a_{\beta_i} + (1-d*A^R_{\beta_i})w_4D_{\beta_i}
\end{align*}

$d$ is a discounting parameter which is based on a measure of the amount of attack surface exposed. $d$ is low in the initial rounds when the defender does not have enough of the right kind of data, but would gradually increase as more information about the attacker's preferences about various regions of the attack surface become available.

Now, let us look at how reasoning about unexposed portions of the attack surface based on the exposed areas. If a target profile $\beta_u$ was not exposed to attacker response in round $r$, the defender will not be able to compute the vulnerability $V^r_{\beta_u}$. Therefore, it is not possible to estimate the attractiveness for $\beta_u$ and hence the optimal defender strategy. So, in keeping with the analysis on available data and based on the spillover effect introduced earlier, the distance--weighted $k$--nearest neighbors algorithm is employed to obtain the vulnerability $V^r_{\beta_u}$ of an unexposed target profile $\beta_u$ in round $r$, based on the $k$ most similar target profiles which were exposed to the attacker in round $r$.

\subsubsection{Generating defender's strategies against SHARP}
While SHARP provides an adversary model, the defender strategies against such model must be generated. To that end, first the parameters of SHARP from available data are learned. Then, future round strategies against the boundedly rational adversary are generated, characterized by the learned model parameters by solving the following optimization problem:

\begin{equation*}
max_{x \in X} \left[\sum_{i \in T} U^d_i(x)q^R_i(w|x) \right]
\end{equation*}

where $q^R_i(w|x)$ is the probability that the adversary will attack target $i$ in round $R$.

\subsection{Learning adversary models from three defender's strategies}
Here, a new approach to learn the parameters of the behavioral model of a bounded rational attacker (thereby pinpointing
a near optimal strategy) is developed, by observing how the attacker responds to only three defender strategies.

\textit{Note: even though the setting is the same of the previous sections, since some assumptions changed, e.g., the adaptivity of the attacker, we slightly change the adopted notation to avoid confusion.}\\

Let the utility function of the attacker for target $t \in T$ be denoted by $u_t : [0, 1] \rightarrow R$. Given a coverage probability vector $\boldsymbol{x} \in \boldsymbol{X}$, the utility of the attacker under strategy $\boldsymbol{x}$ is defined as $u_t(x_t)$. Upon observing the defender's strategy $\boldsymbol{x}$, the attacker computes the utility on each target $t$, $u_t(x_t)$, and based on these utilities responds to the defender's strategy. Here, a non--adaptive attacker is considered. She attacks target $t$ with probability:

\begin{equation*}
D^{\boldsymbol{x}}(t) = \frac{e^{u_t(x_i)}}{\sum_{i \in T} e^{u_i(x_i)}}
\end{equation*}

Our model is a generalization of bounded rationality models such as SUQR. Suppose the same mixed strategy $\boldsymbol{x}$ is played for multiple time steps. The empirical distribution of attacks on target $t$ under $\boldsymbol{x}$ is denoted by $\hat{D}^{\boldsymbol{x}}(\cdot)$. Furthermore, it is assumed that for the strategies, and for all $t$, $D^{\boldsymbol{x}}(t) \geq \rho$, for some $\rho = \frac{1}{poly(n)}$. This assumption is required to estimate the value of $D^{\boldsymbol{x}}(t)$ with polynomially many samples. Our goal is to learn the utility functions, $u_t(\cdot)$ for all $t \in T$, by observing attacker's responses to a choice of coverage probability vectors $\boldsymbol{x} \in \boldsymbol{X}$. This allows to find an approximately optimal defender strategy --- the strategy that leads to the best defender utility. $\hat{u}_t : [0, 1] \rightarrow R$ uniformly approximates or uniformly learns $u_t(\cdot)$ within an error of $\epsilon$, if $\forall x \in [0, 1], |\hat{u}_t(x) - u_t(x)| \leq \epsilon$. Note that the attacker's mixed strategy remains the same when the utility functions corresponding to all targets are increased by the same value. Therefore, only a normalized representation of the utility functions can be learned, such that for all $t$ and all $x$, $|\hat{u}_t(x) + c - u_t(x)| \leq \epsilon$ for some $c$.

\subsubsection{Linear utility functions}
Assume that the utility functions are linear and denoted by $u_t(x) = w_tx + c_t$. Normalizing the utilities, without loss of generality, $c_n = 0$.

\begin{theorem}
Suppose the functions $u_1(\cdot), \ldots, u_n(\cdot)$ are linear. Consider any 3 strategies $\boldsymbol{x}, \boldsymbol{y}, \boldsymbol{z} \in \boldsymbol{X}$ such that for any $t < n, |(x_t-y_t)(x_n-z_n) - (x_n-y_n)(x_t-z_t)| \geq \lambda$, and for any two different strategies $\boldsymbol{p}, \boldsymbol{q} \in \{\boldsymbol{x},\boldsymbol{y},\boldsymbol{z}\}$, it holds $|p_t-q_t| \geq \nu$. If it is possible to access $m = \Omega\left(\frac{1}{\rho}\left(\frac{1}{\epsilon \nu \lambda}\right)^2 log\left(\frac{n}{\delta}\right)\right)$ samples of each of these strategies, then with probability $1-\delta$, each $u_t(\cdot)$ can be uniformly learned within error $\epsilon$.
\end{theorem}

$\nu$ depends on how different the strategies are from each other --- a very small value means that they are almost identical on some coordinates. The lower bound of $\lambda$ would not be very small unless there is a very specific relation between the strategies. As a sanity check, if the three strategies were chosen uniformly
at random from the simplex, both values would be at least $1/poly(n)$. In the quantal best--response model, for each strategy $\boldsymbol{x}$, the ratio between the attack probabilities of two targets $t$ and $n$ follows the relation:

\begin{equation*}
u_t(x_t) = ln\left(\frac{D^{\boldsymbol{x}}(t)}{D^{\boldsymbol{x}}(n)} \right)
\end{equation*}

Therefore, each strategy induces $n-1$ linear equations that can be used to solve for the coefficients of $u_t$. However, only an estimate $\hat{D}^{\boldsymbol{x}}(t)$ of the probability that target $t$ is attacked under a strategy $\boldsymbol{x}$, based on the given samples, can be obtained. So, the inaccuracy in our estimates of $ln\left(\frac{\hat{D}^{\boldsymbol{x}}(t)}{\hat{D}^{\boldsymbol{x}}(n)} \right)$  leads to inaccuracy in the estimated polynomial $\hat{u}_t$. For sufficiently accurate estimates $D^{\boldsymbol{x}}(t)$, the value of $u_t$ differs from the true value by at most $\epsilon$.

\begin{lemma}
Given $\boldsymbol{x} \in \boldsymbol{X}$, let $D^{\boldsymbol{x}}(t)$ be the empirical distribution of attacks based on $m = \Omega \left(\frac{1}{\rho}\left(\frac{1}{\epsilon \nu \lambda}\right)^2 log\left(\frac{n}{\delta}\right)\right)$ samples. With probability $1-\delta$, for all $t \in T$, $\frac{1}{\epsilon} \leq \frac{\hat{D}^{\boldsymbol{x}}(t)}{D^{\boldsymbol{x}}(t)} \leq 1+\epsilon$.
\end{lemma}

\begin{theorem}
Suppose the functions $u_1(\cdot), \ldots, u_n(\cdot)$ are polynomials of degree at most $d$. Consider any $2d + 1$ strategies, $\boldsymbol{y}^{(1)}, \ldots, \boldsymbol{y}^{(d)}, \boldsymbol{y}^{(d+1)} = \boldsymbol{x}^{(1)}, \ldots, \boldsymbol{x}^{(d)} \boldsymbol{x}^{(d+1)}$ such that for all $k, k', k \neq k', y_1^{(k)} = y_1^{(k')}, p_n^{(k)} = p_n^{(k')}, |y_n^{(k)}-y_n^{(k')}| \geq \nu$ and for all $t<n, |x_t^{(k)}-x_t^{(k')}| \geq \nu$. If it is possible to have access to $m = \Omega\left(\frac{1}{\rho}\left(\frac{1}{\epsilon \nu \lambda}\right)^2 log\left(\frac{n}{\delta}\right)\right)$ samples of each of these strategies, then with probability $1-\delta$, each $u_t(\cdot)$ can be uniformly learned within error $\epsilon$.
\end{theorem}

Now, any utility function that is continuous and $L$--Lipschitz, i.e., for all $t$ and values $x$ and $y$, $|u_t(x) - u_t(y)| \leq L|x - y|$, should be learned. Such utility functions can be uniformly learned up to error $\epsilon$, using $O\left(\frac{L}{\epsilon}\right)$ strategies. For any $L$--Lipschitz function $u_t(x)$, there is a
polynomial of degree $m = 12L/\epsilon$ that uniformly approximates
$u_t(x)$ within error of $\epsilon/2$.

\begin{theorem}
Suppose the functions $u_1(\cdot), \ldots, u_n(\cdot)$ are
L-Lipschitz. For $d = 12L/\epsilon$, consider any $2d+1$ strategies, $\boldsymbol{y}^{(1)}, \ldots, \boldsymbol{y}^{(d)}, \boldsymbol{y}^{(d+1)} = \boldsymbol{x}^{(1)}, \ldots, \boldsymbol{x}^{(d)} \boldsymbol{x}^{(d+1)}$ such that for all $k, k', k \neq k', y_1^{(k)} = y_1^{(k')}, p_n^{(k)} = p_n^{(k')}, |y_n^{(k)}-y_n^{(k')}| \geq \nu$ and for all $t<n, |x_t^{(k)}-x_t^{(k')}| \geq \nu$. If it is possible to have access to $m = \Omega\left(\frac{L^2}{\rho \epsilon^{24L/\epsilon}} log\left(\frac{n}{\delta}\right)\right)$ samples of each of these strategies, then with probability $1-\delta$, each $u_t(\cdot)$ can be uniformly learned within error $\epsilon$.
\end{theorem}

\subsubsection{Learning the optimal strategy}
So far, the focus has been on the problem of uniformly learning the utility function of the attacker. Now, it is shown that an accurate estimate of this utility function allows to pinpoint an almost optimal strategy for the defender. Let the utility function of the defender on target $t \in T$ be denoted by $v_t : [0, 1] \rightarrow [-1, 1]$. Given a coverage probability vector $\boldsymbol{x} \in \boldsymbol{X}$, the utility the defender receives when target $t$ is attacked i $v_t(x_t)$. The overall expected utility of the defender is:

\begin{equation*}
V(\boldsymbol{x}) = \sum_{t \in T} D^{\boldsymbol{x}}(t)v_t(x_t)
\end{equation*}

Let $\hat{u}_t$ be the learned attacker utility functions, and $\bar{D}^{\boldsymbol{x}}(t)$ be the predicted attack probability on target $t$ under strategy $\boldsymbol{x}$, according to the utilities $\hat{u}_t$, i.e.:

\begin{equation*}
\bar{D}^{\boldsymbol{x}}(t) = \frac{e^{\hat{u}_t(x_i)}}{\sum_{i \in T} e^{\hat{u}_i(x_i)}}
\end{equation*}

Let $\bar{V}(\boldsymbol{x})$ be the predicted expected utility of the defender based on the learned attacker utilities $\bar{D}^{\boldsymbol{x}}(t)$, i.e., $\bar{V}(\boldsymbol{x}) = \sum_{t \in T} \bar{D}^{\boldsymbol{x}}(t)v_t(x_t)$. When the attacker utilities are uniformly learned within error $\epsilon$, then $\bar{V}$ estimates $V$ with error at most $8\epsilon$. At a high level, this is established by showing that one can predict the attack distribution using the learned attacker utilities. Furthermore, optimizing the defender's strategy against the approximate attack distributions leads to an approximately optimal strategy for the defender.

\begin{theorem}
Assume for all $\boldsymbol{x}$ and any $t \in T$, $|\hat{u}_t(x_t)-u_t(x_t)| \leq \epsilon \leq \ 1/4$. Then, for all $\boldsymbol{x}$, $|\bar{V}(\boldsymbol{x}) - V(\boldsymbol{x})| \leq 4\epsilon$. Furthermore, let $\boldsymbol{x'} = \arg \max_{\boldsymbol{x}} \bar{V}(\boldsymbol{x})$ be the predicted optimal strategy, then $\max_{\boldsymbol{x}} V(\boldsymbol{x}) - V(\boldsymbol{x'}) \leq 8\epsilon.$
\end{theorem}
\section{Determining attacker's payoffs exploiting regret--based solutions}\label{sec:mmr}
The adversary behavior models (capturing bounded rationality) can be learned from real--world data on where adversaries have attacked, and game payoffs can be determined precisely from data on animal densities.

One key approach to modeling payoff uncertainty is to express the adversary's payoffs as lying within specific intervals: for each target $t$, it holds:

\begin{equation*}
R^a_t \in [{R^a}_{min}(t), {R^a}_{max}(t)], P^a_t \in [{P^a}_{min}(t), {P^a}_{max}(t)]
\end{equation*}

While a behavioral model from real--world data can be learned, access to data to precisely compute animal density may be not always possible. For example, given limited numbers of rangers, they may have patrolled and collected wildlife data from only a small portion of a national park, and thus payoffs in other areas of the park may remain uncertain. Also, due to the dynamic changes (e.g., animal migration), players' payoffs may become uncertain in the next season. Hence, a new MiniMaxRegret (MMR)--based robust algorithm is introduced, ARROW, to handle payoff uncertainty in Green Security Games (a.k.a. GSGs, i.e., Security Games for the protection of wildlife, forest and fisheries), taking into account adversary behavioral models. Here, the main focus is on zero--sum games, as motivated by recent work in green security domains and earlier major SSG applications that use zero--sum games. In addition, a model inspired by SUQR is adopted, where the subjective utility function is:

\begin{equation*}
\hat{U}^a_t(\boldsymbol{x}, \boldsymbol{R^a}, \boldsymbol{P^a}) = w_1x_t + w_2R^a_t + w_3P^a_t + w_4\Phi_t
\end{equation*}

where $x_t$ is the coverage probability of target $t$ and $\Phi_t$ is another feature of target $t$, e.g., distance, as seen for SHARP. MMRb with uncertain payoffs is now formulated for both players in zero--sum SSG with a boundedly rational attacker.

\begin{definition}
Given $(\boldsymbol{R^a}, \boldsymbol{P^a})$, the defender's behavioral regret is the loss in her utility for playing a strategy $\boldsymbol{x}$ instead of the optimal strategy, which is represented as follows:

\begin{equation*}
R_b(\boldsymbol{x}, \boldsymbol{R^a}, \boldsymbol{P^a}) = max_{\boldsymbol{x'} \in \boldsymbol{X}} F(\boldsymbol{x'}, \boldsymbol{R^a}, \boldsymbol{P^a}) - F(\boldsymbol{x}, \boldsymbol{R^a}, \boldsymbol{P^a})
\end{equation*}

where

\begin{equation*}
F(\boldsymbol{x}, \boldsymbol{R^a}, \boldsymbol{P^a}) = \sum_t \hat{q}_t(\boldsymbol{x}, \boldsymbol{R^a}, \boldsymbol{P^a})U^d_t(\boldsymbol{x}, \boldsymbol{R^d}, \boldsymbol{P^d})
\end{equation*}
\end{definition}

Behavioral regret measures the distance in terms of utility loss from the defender strategy x to the optimal strategy given the attacker payoffs. Here, $F(\boldsymbol{x}, \boldsymbol{R^a}, \boldsymbol{P^a})$ is the defender's utility for playing $\boldsymbol{x}$ where the attacker payoffs, whose response follows SUQR, are $(\boldsymbol{R^a}, \boldsymbol{P^a})$. The defender's payoffs in zero-sum games are $\boldsymbol{R^d} = -\boldsymbol{P^a}$ and $\boldsymbol{P^d} = -\boldsymbol{R^a}$. When the payoffs are uncertain, if the defender plays a strategy $\boldsymbol{x}$, she receives different behavioral regrets w.r.t to different payoff instances within the uncertainty intervals. Thus, she could receive a behavioral max regret which is defined as follows.

\begin{definition}
Given payoff intervals $I$, the behavioral max regret for the defender to play a strategy $\boldsymbol{x}$ is the maximum behavioral regret over all payoff instances:

\begin{equation*}
MR_b(\boldsymbol{x},\boldsymbol{I}) = max_{(\boldsymbol{R^a}, \boldsymbol{P^a}) \in I} R_b(\boldsymbol{x}, \boldsymbol{R^a}, \boldsymbol{P^a})
\end{equation*}
\end{definition}

\begin{definition}
Given payoff intervals $I$, the behavioral minimax regret problem attempts to find the defender optimal strategy that minimizes the MRb she receives:

\begin{equation*}
MMR_b(\boldsymbol{I}) = min_{\boldsymbol{x} \in \boldsymbol{X}} MR_b(\boldsymbol{x},\boldsymbol{I})
\end{equation*}
\end{definition}

Intuitively, behavorial minimax regret ensures that the defender's strategy minimizes the loss in the solution quality over the uncertainty of all possible payoff realizations.

Overall, MMRb can be reformulated as minimizing the max regret $r$ such
that $r$ is no less than the behavioral regrets over all payoff instances within the intervals:

\begin{align*}
&min_{\boldsymbol{x} \in \boldsymbol{X}, r \in \mathbb{R}} \quad r \\
&s.t. \quad r \geq F(\boldsymbol{x'}, \boldsymbol{R^a}, \boldsymbol{P^a}) - F(\boldsymbol{x}, \boldsymbol{R^a}, \boldsymbol{P^a}), \forall (\boldsymbol{R^a}, \boldsymbol{P^a}) \in I, \boldsymbol{x'} \in \boldsymbol{X} 
\end{align*}

The set of constraints is infinite since $X$ and $I$ are continuous. One practical approach to optimization with large constraint sets is constraint sampling, coupled with constraint generation (a.k.a. row generation). Following this approach, ARROW samples a subset of constraints and gradually expands this set by adding violated constraints to the relaxed problem until convergence to the optimal $MMR_b$ solution.

Specifically, ARROW begins by sampling pairs $(\boldsymbol{R^a}, \boldsymbol{P^a})$ of the adversary payoffs uniformly from $I$. The corresponding optimal strategies for the defender given these payoff samples, denoted $x'$, are then computed using the PASAQ algorithm~\cite{yang2012computing} to obtain a finite set $S$ of sampled constraints. These sampled constraints are then used to solve the corresponding relaxed MMRb program using the R.ARROW algorithm. The optimal solution $(lb,\boldsymbol{x^*})$ is thus obtained, providing a lower bound ($lb$) on the true $MMR_b$. Then constraint generation is applied to determine violated constraints (if any). This uses the M.ARROW algorithm which computes $MR_b(\boldsymbol{x^*}, \boldsymbol{I})$, the optimal regret of $\boldsymbol{x^*}$ which is an upper bound ($ub$) on the true MMRb. If $ub > lb$, the optimal solution of M.ARROW, $\{\boldsymbol{x^{',*}}, \boldsymbol{R^{a,*}}, \boldsymbol{P^{a,*}}\}$ provides the maximally violated constraint, which is added to $S$. Otherwise, $\boldsymbol{x^*}$ is the minimax optimal strategy and $lb = ub = MMR_b(\boldsymbol{I})$.

The first step of ARROW is to solve the relaxed $MMR_b$ problem using R.ARROW. This relaxed $MMR_b$ problem is non--convex. Thus, R.ARROW presents two key ideas for efficiency:

\begin{enumerate}
\item binary search (which iteratively searches the defender's utility space to find the optimal solution) to remove the fractional terms in relaxed $MMR_b$;
\item it then applies piecewise--linear approximation to linearize the non--convex terms of the resulting decision problem at each binary search step.
\end{enumerate}

\begin{theorem}
R.ARROW provides an $O\left(\epsilon + \frac{1}{M} \right)$--optimal solution of relaxed $MMR_b$ where $\epsilon$ is the tolerance of binary search and $M$ is the number of piecewise segments.
\end{theorem}

Given the optimal solution $\boldsymbol{x^*}$ returned by R.ARROW, the second step of ARROW is to compute $MR_b$ of $\boldsymbol{x^*}$ using M.ARROW. A local search with multiple starting points is employed since allows to reach different local optima.

While ARROW incorporates an adversary behavioral model, it may not be applicable for green security domains where there may be a further paucity of data in which not only payoffs are uncertain but also parameters of the behavioral model are difficult to learn accurately. Therefore, a novel MMR-based algorithm, ARROW--Perfect, can be introduced to handle uncertainty in both players' payoffs assuming a perfectly rational attacker. In general, ARROW--Perfect follows the same constraint sampling and constraint generation methodology as ARROW. Yet, by leveraging the property that the attacker's optimal response is a pure strategy (given a perfectly rational attacker) and the game is zero--sum, the exact optimal solutions for computing both relaxed MMR and max regret in polynomial time are obtained (while we cannot provide such guarantees for a boundedly rational attacker).
\section{Online learning}\label{sec:mab}

\subsection{Handling exploration--exploitation tradeoffs in Security Games}
Previous research optimizes defenders' strategies by modeling this problem as a repeated Stackelberg game, capturing the special property in this domain frequent interactions between defenders and attackers. However, this research fails to handle exploration--exploitation tradeoff in this domain caused by the fact that defenders only have knowledge of attack activities at targets they protect. The problem is formulated as a Restless Multi--Armed Bandit (RMAB) model to address this challenge. To use Whittle index policy to plan for patrol strategies in the RMAB, two sufficient conditions for indexability and an algorithm to numerically evaluate indexability are provided. Given indexability, a binary search based algorithm to find Whittle index policy efficiently is proposed.

It is assumed that defenders have knowledge of all poaching activities throughout the wildlife protected area. Unfortunately, given vast geographic areas for wildlife protection, defenders do not have knowledge of poaching activities in areas they do not protect. Thus, defenders are faced with the exploration--exploitation tradeoff. The exploration--exploitation tradeoff here is different from that in the non--Bayesian stochastic multi--armed bandit problem. In stochastic multi--armed bandit problems, the rewards of every arm are random variables with a stationary unknown distribution. However, in our problem, patrol affects attack activities, i.e.m, more patrol is likely to decrease attack activities and less patrol is likely to increase attack activities. Thus, the random variable distribution is changing depending on player's choice --- more selection (patrol)
leads to lower reward (less attack activities) and less selection (patrol) leads to higher reward (more attack activities). On the other hand, adversarial multi--armed bandit problem is also not an appropriate model for this domain. In adversarial multi--armed bandit problems, the reward can arbitrarily change while the attack activities
in our problem are unlikely to change rapidly in a short period.

Poaching activity is a dynamic process affected by patrol. If patrollers patrol in a certain location frequently, it is very likely that the poachers poaching in this location will switch to other locations for poaching. On the other hand, if a location has not been patrolled for a long time, poachers may gradually notice that and switch to this location for poaching. In the wildlife protection domain, both patrollers and poachers do not have perfect observation of their opponents' actions. This observation imperfection lies in two aspects: limited observability --- patrollers/poachers do not know what happens at locations they do not patrol/poach; partial observability --- patrollers/poachers do not have perfect observation even at locations they patrol/poach --- the location might be large (e.g., a $2$km $\times$ $2$km area) so that it is possible that patrollers and poachers do not see each other even if they are at the same location. These two properties make it extremely difficult for defenders to
optimally plan their patrol strategies.	

\subsubsection{Problem formulation}
There are $n$ targets that are indexed by $N = \{1, \ldots, n\}$. Defenders have $k$ patrol resources that can be deployed to these $n$ targets. At every round, defenders choose $k$ targets to protect. After that, defenders will have an observation of the number of attack activities for targets they protect, and no information for targets
they do not protect. The objective for defenders is to decide which $k$ targets to protect at every round to catch as many attackers as possible. Due to the partial observability on defenders' side (defenders' observation of attack activities is not perfect even for targets they protect), a hidden variable attack intensity is introduced, representing the true degree of attack intensity at a certain target. Clearly, this hidden variable attack intensity cannot directly be observed by defenders. Instead, defenders' observation is a random variable conditioned on this hidden variable attack intensity, and the larger the attack intensity is, the more likely it is for defenders to observe more attack activities during their patrol. The hidden variable attack intensity is discretized into $n_s$ levels, denoted by $S = \{0, 1, \ldots, n_s-1\}$. Lower $i$ represents lower attack intensity. For a certain target, its attack intensity transitions after every round. If this target is protected, attack intensity transitions according to a $n_s \times n_s$ transition matrix $T^1$; if this target is not protected, attack intensity transitions according to another $n_s \times n_s$ transition matrix $T^0$. The transition matrix represents how patrol affects attack intensity --- $T^1$ tends to reduce attack intensity and $T^0$ tends to increase attack intensity. Note that different targets may have different transition matrices because some targets may be more attractive to attackers (for example, some locations may have more animal resources in the wildlife protection domain) so that it is more difficult for attack intensity to go down and easier for attack intensity to go up. Also defenders' observations of attack activities are discretized into $n_o$ levels, denoted by $\boldsymbol{O} = \{0, 1, \ldots, n_o-1\}$. Lower $i$ represents less attack activities defenders observe. Note that defenders will only have observation for targets they protect. A $n_s \times n_o$ observation matrix $O$ determines how the observation depends on the hidden variable attack intensity. Generally, the larger the attack intensity is, the more likely it is for defenders to observe more attack activities during their patrol. Similar to transition matrices, different targets may have different observation matrices. While defenders get observations of attack activities during their patrol, they also receive rewards for that --- arresting poachers/fareevaders/smugglers bring benefit. Clearly, the reward defenders receive depends on their observation and thus the reward function is defined as $R(o), o \in \boldsymbol{O}$ --- larger $i$ leads to higher reward $R(i)$. Note that defenders only get rewards for targets they protect. To summarize, for the targets defenders protect, defenders get an observation depending on its current attack intensity, get the reward associated with the observation, and then the attack intensity transitions according to $T^1$; for the targets defenders do not protect, defenders do not have any observation, get reward 0 and the attack intensity transitions according to $T^0$.

\subsubsection{Learning model from defenders' previous observations}
Given defenders' action history $\{a_i\}$ and observation history $\{o_i\}$, our objective is to learn the transition matrices $T^1$ and $T^0$, observation matrix $O$ and initial belief $\pi$. Due to the existence of hidden variables $\{s_i\}$, expectation--maximization (EM) algorithm is used for learning. The update steps are the following:

\begin{align*}
& \pi_i^{(d+1)} = P(s_1 = i|x;\theta^d)\\
& T_{ij}^{1(d+1)} = \frac{\sum_{t=1:a_t=1}^{T-1} P(s_t = i, s_{t+1} = j|x;\theta^d)}{\sum_{t=1:a_t=1}^{T-1} P(s_t = i|x;\theta^d)}\\
& T_{ij}^{0(d+1)} = \frac{\sum_{t=1:a_t=0}^{T-1} P(s_t = i, s_{t+1} = j|x;\theta^d)}{\sum_{t=1:a_t=0}^{T-1} P(s_t = i|x;\theta^d)}\\
& O_{ij}^{(d+1)} = \frac{\sum_{t=1:a_t=1}^{T} P(s_t = i|x;\theta^d)I(o_t = j)}{\sum_{t=1:a_t=1}^{T} P(s_t = i|x;\theta^d)}
\end{align*}

\subsubsection{Restless multi--armed bandit problems}
In RMABs, each arm represents an independent Markov machine. At every round, the player chooses $k$ out of $n$ arms ($k < n$) to activate and receives the reward determined by the state of the activated arms. After that, the states of all arms will transition to new states according to certain Markov transition probabilities. The problem is called \textit{restless} because the states of passive arms will also transition like active arms. The aim of the player is to maximize his cumulative reward by choosing which arms to activate at every round. It  is PSPACE--hard to find the optimal strategy to general RMABs~\cite{papadimitriou1999complexity}. An index policy assigns an index to each state of each arm to measure how rewarding it is to activate an arm at a particular state. At every round, the index policy chooses to pick the $k$ arms whose current states have the highest indices. Since the index of an arm only depends on the properties of this arm, index policy reduces an $n$--dimensional problem to $n$ $1$--dimensional problems so that the complexity is reduced from exponential with $n$ to linear with $n$. Whittle proposed a heuristic index policy for RMABs by considering the Lagrangian relaxation of the problem~\cite{whittle1988restless}. It has been shown that Whittle index policy is asymptotically optimal under certain conditions as $k$ and $n$ tend to infinity with $k/n$ fixed. When $k$ and $n$ are finite, extensive empirical studies have also demonstrated the near--optimal performance of Whittle index policy. Whittle index measures how attractive it is to activate an arm based on the concept of subsidy for passivity. It gives the subsidy $m$ to passive action (not activate) and the smallest $m$ that would make passive action optimal for the current state is defined to be the Whittle index for this arm at this state. Whittle index policy chooses to activate the $k$ arms with the highest Whittle indices. Intuitively, the larger the $m$ is, the larger the gap is between active action (activate) and passive action, the more attractive it is for the player to activate this arm.

Mathematically, denote $V_m(x; a = 0)$ $(V_m(x; a = 1))$ to be the maximum cumulative reward the player can achieve until the end if he takes passive (active) action at the first round at the state $x$ with subsidy $m$. Whittle index $I(x)$ of state $x$ is then defined to be:

\begin{align*}
I(x) = inf_m \{m:V_m(x; a = 0) \geq V_m(x; a = 1)\}
\end{align*}

However, Whittle index only exists and Whittle index policy can only be used when the problem satisfies a property known as indexability. Define $\Phi (m)$ to be the set of states for which passive action is the optimal action given subsidy $m$:

\begin{align*}
\Phi (m) = \{x:V_m(x; a = 0) \geq V_m(x; a = 1)\}
\end{align*}

\begin{definition}
An arm is indexable if $\Phi (m)$ monotonically increases from $\emptyset$ to the whole state space as $m$ increases from $-\infty$ to $+\infty$. An RMAB is indexable if every arm is indexable.
\end{definition}

\subsubsection{Restless bandit formulation}
Every target is viewed as an arm and defenders choose $k$ arms to activate ($k$ targets to protect) at every round. Consider a single arm (target), it is associated with $n_s$ (hidden) states, $n_o$ observations, $n_s \times n_s$ transition matrices $T^1$ and $T^0$, $n_s \times n_o$ observation matrix $O$ and reward function $R(o), o \in O$. For the arm defenders activate, defenders get an observation, get reward associated with the observation, and the state transitions according to $T^1$. Note that defenders' observation is not the state. Instead, it is a random variable conditioned on the state, and reveals some information about the state. For the arms defenders do not activate, defenders do not have any observation, get reward 0 and the state transitions according to $T^0$. Since defenders can not directly observe the state, defenders maintain a belief $b$ of the states for each target, based on which defenders make decisions. The belief is updated according to the Bayesian rules. The following equation shows the belief update when defenders protect this target $(a = 1)$ and get observation $o$ or defenders do not protect this target $(a = 0)$.

\begin{equation*}
b'(s') = \begin{cases} \eta \sum_{s \in \boldsymbol{S}}b(s)O_{so}T_{ss'}^1,& a=1\\
\sum_{s \in \boldsymbol{S}}b(s)O_{so}T_{ss'}^0,& a=0 \end{cases}
\end{equation*}

where $\eta$ is the normalization factor. When defenders do not protect this target $(a = 0)$, defenders do not have any observation, so their belief is updated according to the state transition rule. When defenders protect this target $(a = 1)$, their belief is firstly updated according to their observation $o$ ($b_{new}(s) = \eta b(s)O_{so}$ according to Bayes' rule), and then the new belief is then updated according to the state transition rule:

\begin{align*}
b'(s') & = \sum_{s \in \boldsymbol{S}}b_{new}(s)T_{ss'}^1 \\
& = \sum_{s \in \boldsymbol{S}} \eta b(s)O_{so}T_{ss'}^1 \\
& = \eta \sum_{s \in \boldsymbol{S}} b(s)O_{so}T_{ss'}^1
\end{align*}

Now the mathematical definition of Whittle index is presented. Denote $V_m(b)$ to be the value function for belief state $b$ with subsidy $m$. $V_m(b; a = 0)$ to be the value function for belief state $b$ with subsidy $m$ and defenders take passive action. $V_m(b; a = 1)$ to be the value function for belief state $b$ with subsidy $m$ and defenders take active action. The following equations show these value functions:

\begin{align*}
V_m(b;a = 0) &= m + \beta V_m(b_{a=0})\\
V_m(b;a = 1) &= \sum_{s \in \boldsymbol{S}} b(s) \sum_{o \in \boldsymbol{O}} O_{so}R(o) + \beta \sum_{o \in \boldsymbol{O}} \sum_{s \in \boldsymbol{S}} b(s)O_{so}V_m(b_{a=1}^o)\\
V_m(b) &= max\{V_m(b;a = 0),V_m(b;a = 1)\}
\end{align*}

When defenders take passive action, they get the immediate reward $m$ and the $\beta$--discounted future reward --- value function at new belief $b_a=0$, which is updated from $b$ according to the case $a = 0$. When defenders take active action, they get the expected immediate reward $\sum_{s \in \boldsymbol{S}} b(s) \sum_{o \in \boldsymbol{O}} O_{so}R(o)$ and the $\beta$--discounted future reward. The value function $V_m(b)$ is the maximum of $V_m(b; a = 0)$ and $V_m(b; a = 1)$. Whittle index $I(b)$ of belief state $b$ is then defined to be:

\begin{equation*}
I(b) = inf_m\{m:V_m(b;a = 0) \geq V_m(b;a = 1)\}
\end{equation*}

The passive action set $\Phi(m)$, which is the set of belief states for
which passive action is the optimal action given subsidy $m$ is then
defined to be:

\begin{equation*}
\Phi(m) = inf_m\{b:V_m(b;a = 0) \geq V_m(b;a = 1)\}
\end{equation*}

\subsubsection{Sufficient condition for indexability}
Two sufficient conditions for indexability when $n_o = 2$ and $n_s = 2$ are provided. Denote the transition matrices to be $T^0$ and $T^1$, observation matrix to be $O$. Clearly in our problem, $O_{11} > O_{01}, O_{00} > O_{10}$ (higher attack intensity leads to higher probability to see attack activities when patrolling); $T_{11}^1 > T_{01}^1$, $T_{00}^1 > T_{10}^1$, $T_{11}^0 > T_{01}^0$, $T_{00}^0 > T_{10}^0$ (positively correlated arms).

Define $\alpha = max\{T_{11}^0 - T_{01}^1,T_{11}^0 > T_{01}^1\}$. Since it is a two--state problem with $S = \{0, 1\}$, $x$ represents the belief state: $x = b(s = 1)$, which is the probability of being in state 1.

Define $\Gamma_1(x) = xT_{11}^1 + (1-x)T_{01}^1$ which is the belief for the next round if the belief for the current round is $x$ and the active action is taken. Similarly, $\Gamma_0(x) = xT_{11}^0 + (1-x)T_{01}^0$, which is the belief for the next round if the belief for the current round is $x$ and the passive action is taken.
Below two theorems demonstrating two sufficient conditions for indexability are presented.

\begin{theorem}
When $\alpha \leq 0.5$, the process is indexable, i.e., for
any belief $x$, if $V_m(x; a = 0) \geq V_m(x; a = 1)$, then $V_m′ (x; a =0) \geq Vm′ (x; a = 1), \forall m′ \geq m$.
\end{theorem}

\begin{theorem}
When $\alpha \beta \leq 0.5$ and $\Gamma_1(1) \leq \Gamma_0(0)$, the process is indexable, i.e.,  for any belief $x$, if $V_m(x; a = 0) \geq V_m(x; a = 1)$, then $V_m′ (x; a =0) \geq Vm′ (x; a = 1), \forall m′ \geq m$.
\end{theorem}

\subsubsection{Numerical evaluation of indexability}
\begin{proposition}
If $m < R(0) − \beta \frac{R(n_o-1)-R(0)}{1-\beta}, \Phi(m) = \emptyset$; if $m > R(n_o-1), \Phi(m)$ is the whole belief state space.
\end{proposition}

Thus, it should be determined whether the set $\Phi(m)$ monotonically
increases for $m \subseteq \left[R(0) − \beta \frac{R(n_o-1)-R(0)}{1-\beta}, R(n_o-1) \right]$. Numerically, this limited $m$ range is discretized and then it is evaluated if $\Phi(m)$ monotonically increases with the increase of discretized $m$. Given the subsidy $m$, $\Phi(m)$ can be determined by solving a special POMDP model whose conditional observation probability is dependent on start state and action. The algorithm returns a set $D$ which contains $n_s$--length vectors $d_1, d_2, \ldots, d_{|D|}$. Every vector $d_i$ is associated with an optimal action $e_i$. Given the belief $b$, the optimal action is determined by $a^{opt} = e_i, i = \arg \max_j b^T d_j$. Thus, $\Phi(m) = \cup_{i:e_i=0} \{b:b^Td_i \geq b^Td_j, \forall j\}$.

Given $m_0 < m_1$, our aim is to check whether $\Phi(m_0) \subseteq \Phi(m_0)$. A MILP to verify whether such condition holds can be solved.

\subsubsection{Computation of Whittle index Policy}
Given the indexability, Whittle index can be found by doing a binary
search within the range $m \subseteq \left[R(0) − \beta \frac{R(n_o-1)-R(0)}{1-\beta}, R(n_o-1) \right]$. Given the upper bound $ub$ and lower bound $lb$, the problem with middle point $\frac{lb+ub}{2}$ as passive subsidy is sent to the special POMDP solver to find the optimal action for the current belief. If the optimal action is active, then the Whittle index is greater than the middle point so $lb \leftarrow \frac{lb+ub}{2}$ or else $ub \leftarrow \frac{lb+ub}{2}$. This binary
search algorithm can find Whittle index with arbitrary precision. Naively, the Whittle index policy can be found by computing the $\epsilon$--precision indices of all arms and then picking the $k$ arms with the highest indices.

Specifically, let $A$ be the Whittle index policy to be returned: it is
set to be $\emptyset$ at the beginning. $S$ is the set of arms that are not known whether belonging to $A$ or not and is set to be the whole set of arms at the beginning. Before it finds top--$k$ arms, it tests all the arms in $S$ about their optimal action with subsidy $\frac{lb+ub}{2}$. If the optimal action is 1, it means this arm's index is higher than $\frac{lb+ub}{2}$ and it is added to $S_1$; if the optimal
action is 0, it means this arm's index is lower than $\frac{lb+ub}{2}$ and it is added to $S_0$. At this moment, all arms in $S_1$ have higher indices than all arms in $S_0$. If there is enough space in $A$ to include all arms in $S_1$, $S_1$ is added to $A$, remove them from $S$ and set the upper bound to be $\frac{lb+ub}{2}$ because $S_1$ belongs to Whittle index policy set and all the rest arms have the index lower than $\frac{lb+ub}{2}$. If there is not enough space in $A$, $S_0$ is removed from $S$ and set the lower bound to be $\frac{lb+ub}{2}$ because $S_0$ does not belong to Whittle index policy set and all the rest arms have the index higher than $\frac{lb+ub}{2}$.

\subsubsection{Special POMDP formulation}
Here, the algorithm to compute the passive action set $\Phi(m)$ with the subsidy $m$ is discussed. This problem can be viewed as solving a special POMDP model whose conditional observation probability is dependent on start state and action while the conditional observation probability is dependent on end state and action in standard POMDPs.

The original state is $s$, the agent takes action $a$, and the state transitions to $s'$ according to $P(s'|s, a)$. However, the observation $o$ the agent gets during this process is dependent on $s$ and $a$ in our special POMDPs; while it depends on $s'$ and a in standard POMDPs.

The special POMDP formulation for our problem is straightforward.

\begin{itemize}
\item The state space is $\boldsymbol{S} = \{0, 1, \ldots, n_s-1\}$.
\item The action space is $\boldsymbol{A} = \{0, 1\}$, where $a = 0$ represents passive action (do not protect) and $a = 1$ represents active action (protect).
\item The observation space is $\boldsymbol{O} = \{-1, 0, 1, \ldots, n_o - 1\}$. It adds a \textit{fake} observation $o = -1$ to represent no observation when taking action $a = 0$. It's called fake because defenders have probability 1 to observe o = −1 no matter what the state is when they take action a = 0, so this observation does not provide
any information. When defenders take action $a = 1$, they may
observe observations $\boldsymbol{O} \backslash \{ -1\}$.
\item The conditional transition probability $P(s'|s, a)$ is defined to be $P(s' = j|s = i, a = 1) = T_{ij}^1$ and $P(s' = j|s = i, a = 0) = T_{ij}^0$.
\item The conditional observation probability $P(o|s, a)$ is defined to be $P(o =-1|s, a = 0) = 1, \forall s \in S; P(o = j|s = i, a = 1) = \boldsymbol{O}_{ij}$. Note that the conditional observation probability here is dependent on the start state $s$ and action $a$, while it depends on end state $s′$ and action $a$ in standard POMDP models. Intuitively, defenders' observation of attack activities today depends on the attack intensity today, not the transitioned attack intensity tomorrow.
\item The reward function R is:
\begin{equation*}
R(s,s',a,o) = \begin{cases} 0,&a=0\\R(o),&a=1 \end{cases}
\end{equation*}
\end{itemize}

With the transition probability and observation probability, $R(s, a)$ can be computed. Note that this formulation is also slightly different
due to the different definition of observation probability.

\begin{equation*}
R(s,a) = \sum_{s' \in \boldsymbol{S}} P(s'|s,a) \sum_{o \in \boldsymbol{O}} P(o|s,a)R(s,s',a,o)
\end{equation*}

\subsubsection{Value iteration for the special POMDP}
Different from standard POMDP formulation, the belief update
in the special POMDP formulation is:

\begin{equation*}
b'(s') = \frac{\sum_{s' \in \boldsymbol{S}}b(s)P(o|s,a)P(s'|s,a)}{P(o|b,a)}
\end{equation*}

where

\begin{equation*}
P(o|b,a) = \sum_{s' \in \boldsymbol{S}} \sum_{s \in \boldsymbol{S}} b(s)P(o|s,a)P(s'|s,a) = \sum_{s \in \boldsymbol{S}}b(s)P(o|s,a)
\end{equation*}

Similar to standard POMDP formulation, the value function is:

\begin{equation*}
V'(b) = max_{a \in \boldsymbol{A}} \left(\sum_{s \in \boldsymbol{S}} b(s)R(s,a) + \beta \sum_{o \in \boldsymbol{O}} P(o|b,a)V(b_a^o) \right)
\end{equation*}

which can be broken up to simpler combinations of other value functions:

\begin{align*}
	&V'(b) = max_{a \in \boldsymbol{A}} V_a(b) \\
	&V_a(b) = \sum_{o \in \boldsymbol{O}} V_a^o(b) \\
	&V_a^o(b) = \frac{\sum_{s \in \boldsymbol{S}} b(s)R(s,a)}{|\boldsymbol{O}|} + \beta P(o|b,a)V(b_a^o)
\end{align*}

All the value functions can be represented as $V(b) = max_{\alpha \in \boldsymbol{D}} b\alpha$ since the update process maintains this property, so the set $D$ is updated when updating the value function.

\subsubsection{Planning from POMDP view}
Every single target can be modeled as a special POMDP model. Given that, these POMDP models at all targets can be combined to form a special POMDP model that describe the whole problem, and solving this special POMDP model leads to defenders' exact optimal strategy. Use the superscript $i$ to denote target $i$. Generally, the POMDP model for the whole problem is the cross product of the single--target POMDP models at all targets with the constraint that only $k$ targets are protected at every round.

\begin{itemize}
\item The state space is $\boldsymbol{S} = \boldsymbol{S^1} \times \boldsymbol{S^2} \times \ldots \times \boldsymbol{S^n}$. Denote
$s = (s^1, s^2, \ldots, s^n)$.
\item The action space is $\boldsymbol{A} = \{(a^1, a^2, \ldots, a^n)|a^j \in \{0, 1\}, \forall j \in \mathbb{N}, \sum_{j \in \mathbb{N}} a^j = k\}$, which represents that only $k$ targets can be protected at a round. Denote $a = (a^1, a^2, \ldots, a^n)$.
\item The observation space $\boldsymbol{O} = \boldsymbol{O^1} \times \boldsymbol{O^2} \times \ldots \times \boldsymbol{O^n}$. Denote
$o = (o^1, o^2, \ldots, o^n)$.
\item The conditional transition probability is $P(s'|s, a) = \prod_{j \in \mathbb{N}} P^j (s^{'j} |s^j , a^j)$.
\item The conditional observation probability is $P(o|s, a) = \prod_{j \in \mathbb{N}} P^j (o^{j} |s^j , a^j)$.
\item The reward function is $R(s, s', a, o) = \prod_{j \in \mathbb{N}} R(s^j,s^{'j,a^j,o^j})$.
\end{itemize}

Silver and Veness~\cite{silver2010monte} have proposed POMCP algorithm, which provides high quality solutions and is scalable to large POMDPs. The POMCP algorithm only requires a simulator of the problem so it also applies to our special POMDPs. At a high level, the POMCP algorithm is composed of two parts: it uses a particle filter to maintain an approximation of the belief state; it draw state samples from the particle filter and then use MCTS to simulate what will happen next to find the best action. It uses a particle filter to approximate the belief state because it is even computationally impossible in many problems to update belief state due to the extreme large size of the state space. However, in our problem, the all--target POMDP model is the cross product of the single--target POMDP models at all targets. The single--state POMDP model is small so that it is computationally inexpensive to maintain its belief state. Thus, the state $s^i$ at target $i$ can be sampled from its belief state and then compose them together to get the state sample $s = (s^1, s^2, \ldots, s^n)$ for the all--target POMDP model.

\subsection{Online planning for optimal protector strategies in resource conservation games}
Protectors (law enforcement agencies) try to protect natural resources, while extractors (criminals) seek to exploit them. In many domains, such as illegal fishing, the extractors know more about the distribution and richness of the resources than the protectors, making it extremely difficult for the protectors to optimally allocate their assets for patrol and interdiction. Fortunately, extractors carry out frequent illegal extractions, so protectors can learn about the richness of resources by observing the extractor's behavior.

In resource conservation domains, the protector often does not know the distribution of resources while the extractor may have more information about it, e.g. preventing illegal fishing. Our goal is to provide an optimal asset deployment (e.g., patrol) strategy for the protector, given her lack of knowledge about the distribution of resources.

\subsubsection{Problem formulation}
In resource conservation games, the extractor's frequent illegal extractions provide the protector with the opportunity to learn about the distribution of resources by observing the extractor's behavior. The aim is the construction of an online policy for the protector to maximize her utility given observations of the extractor. At every round, the protector chooses one site to protect and the extractor simultaneously chooses one site to steal from. Both the extractor and the protector have full knowledge about each other's previous actions.

In our model, the amount of resources at each site will be fixed and the extractor will have full knowledge of this distribution. The protector will have to learn this distribution by observing the extractor's behavior.

There is a finite time horizon $t \in T$. There are $n$ sites indexed by $N = \{1, 2, \ldots, n\}$ that represent the locations of the natural resource in question: the extractor wants to steal resources from these sites and the protector wants to interdict the extractor. The value of the sites to the extractor is represented in terms of their utilities. Each site has a utility $u(i)$ that is only known to the extractor. The utility space is discretized into $m$ levels, $u(i) \in M = \{1, 2, \ldots ,m\}$. Human beings cannot distinguish between tiny differences in utilities in the real world, so discretizing these utilities is justified. For $n$ sites and $m$ utility levels, there are $m^n$ possible sets of utilities across all sites. The distribution of resources is then captured by the vector of utilities at each site, and the set of possible resource distributions is:

\begin{equation*}
U = \{(u(1),u(2), \ldots, u(n)):u(i) \in M, \forall i \in N\}
\end{equation*}

Assuming that the resource levels $u(i)$, $i \in N$ are independent from each other, at the beginning of the game, the protector may have some prior knowledge about the resource levels $u(i)$ at each site $i \in N$. This prior knowledge is represented as a probability density function $p(u(i))$ over $M$. If the protector does not know
anything about $u(i)$, then a uniform prior for $u(i)$ over $M$ is adopted. At each time $t \in T$, the protector chooses a site $a_t \in N$ to protect and the extractor simultaneously chooses a site $o_t \in N$ from which to steal. If $a_t = o_t$, the protector catches the extractor and the extractor is penalized by the amount $P(o_t) < 0$; if $a_t \neq o_t$, the extractor successfully steals resources from site $o_t$ and gets a payoff of $u(o_t)$. For clarity, the protector's interdiction is always successful whenever it visits the same site as the extractor. Additionally, the protector fully observes the moves of the extractor, likewise, the extractor fully observes the moves of the protector. Note that the penalty $P(i), i \in N$ is known to both the protector and the extractor. A zero--sum game is adopted, so the protector is trying to minimize the extractor's payoffs. In most resource conservation domains, the extractor pays the same penalty $P$ if he is seized independent of the site he visits. Varying penalties across sites for greater generality are allowed. A fictitious Quantal Response playing (FQR) extractor is assumed. Specifically, a fictitious extractor assumes the protector's empirical distribution will be his mixed strategy in the next round. In this behavior model, the extractor makes decisions based on the parameters $u(i), P(i), i \in N$, as well as the protector's actions in previous rounds.

The extractor behaves in the following way: in every round, he computes the empirical coverage probability $c_i$ for every site $i$ based on the history of the protector's actions, then computes the expected utility $EU(i) = c(i)P(i) + (1-c(i))u(i)$ for every site and finally attempts to steal from the site i with the probability proportional to $e^{\lambda EU(i)}$ where $\lambda \geq 0$ is the parameter representing the rationality of the player (higher $\lambda$ represents a more rational player).

To implement the above model, two technical questions must be resolved. First, at every round $t$, based on her current belief about $u$, how should the protector choose sites to protect in the next round? Second, after each round, how should the protector use the observation of the latest round to update her beliefs about $u$? Here, decision making and belief updating in a partially observable environment, where the payoffs $u$ are unobservable and the extractor's actions are observable, are being studied, which is the exact setup for a POMDP. A two--player game is now setup as a POMDP $\{S,A,O,T,\Omega,R\}$ where the extractor follows a quantal response model.

\begin{itemize}
\item The state space of our POMDP is $S = U \times Z^n$, which
is the cross product of the utility space and the count space. $U$ is
the utility space. $Z^n$ is the set of possible counts of the protector's visits to each site, where $C_t \in Z^n$ is an
integer-valued vector where $C_t(i) , i \in N$ is the number of times
that the protector has protected site $i$ at the beginning of round $t \in T$. A particular state $s \in S$ is written as $s = (u,C)$, where $u$ is the vector of utility levels for each site and $C$ is the current state count. The initial beliefs are expressed by a distribution over $s = (u, 0)$, induced by the prior distribution on $u$. $c_t(i)$ = $\frac{C_t(i)}{t-1}$ denotes the frequency with which the protector visits site $i$ at the beginning of round $t \in T$. $c_1 = 0$ by convention.
\item The action space $A$ is $N$, representing the site the protector chooses to protect.
\item The observation space $O$ is $N$, representing the site the extractor chooses to attempt to steal from.
\item Let $e_a \in R^n$ denote the unit vector with a 1 in slot $a \in N$ and zeros elsewhere. The conditional transition probability $T$ governing the evolution of the state is:

\begin{equation*}
T(s' = (u',C')| s = (u,C),a)=\begin{cases} 1 & u=u', C' = C+e_a,\\
0, & otherwise \end{cases}
\end{equation*}

Specifically, the evolution of the state is deterministic. The underlying utilities do not change, and the count for the site visited by the protector increases by one while all others stay the same.
\item $EU(u,C) \in R^n$ defines the vector of empirical expected utilities for the extractor for all sites when the actual utility is $u$ and the count is $C$,

\begin{equation*}
[EU(u, C)](i) = c(i)P(i) + (1-c(i))u(i), \forall i \in N,
\end{equation*}

when $t \geq 1$. $[EU (u, 0)](i) = u(i)$ by convention. Hence,
our observation probabilities are explicitly:

\begin{equation*}
\Omega(o|s'=(u,C),a)=\frac{e^{\lambda [EU(u,C-e_a)](o)}}{\sum_{i \in N} e^{\lambda [EU(u,C-e_a)](i)}}
\end{equation*}

the probability of observing the extractor takes action $o$ when the
protector takes action $a$ and arrives at state $s′$. Note that both $a$ and $o$ are the actions the protector/extractor take at the same round.
\item The reward function $R$ is:

\begin{equation*}
R(s=(u,C),s'=(u,C+e_a),a,o) = \begin{cases}-P(o), & a = o\\
-u(o), & a \neq o \end{cases}
\end{equation*}
\end{itemize}

\subsubsection{GMOP algorithm}
The size of the utility space $U$ is $m^n$, and the size of the count space is $O\left(\frac{T^n}{n!}\right)$. The computational cost of the latest POMDP solvers soon become unaffordable for us as the problem size grows.

Silver and Veness~\cite{silver2010monte} have proposed the POMCP algorithm, which provides high quality solutions for large POMDPs. The POMCP algorithm uses a particle filter to approximate the belief state. Then, it uses Monte Carlo tree search (MCTS) for online planning where state samples are drawn from the particle filter and the action with the highest expected utility based on Monte Carlo simulations is chosen. However, the particle filter is only an approximation of the true belief state and is likely to move further away from the actual belief state as the game goes on, especially when most particles get depleted and new particles need to be added. Adding new particles will either make the particle filter a worse approximation of the exact belief state, if the added particles do not follow the distribution of the belief state or be as difficult as drawing samples directly from the belief state, if the added particles do follow the distribution of the belief state.

Our POMDP has specific structure that can be exploited. The count state in $S$ is known and the utility state does not change, making it possible to draw samples directly from the exact belief state using Gibbs sampling. We propose the GMOP algorithm that draws samples directly from the exact belief state using Gibbs sampling, and then runs MCTS. The samples drawn directly from the belief state better represent the true belief state compared to samples drawn from a particle filter.

At a high level, in round $t$ the protector draws samples of state $s$ from its belief state $B_t(s)$ using Gibbs sampling and then it runs MCTS using those samples. Finally, it executes the action with the highest expected utility. MCTS starts with a tree that only contains a root node. Since the count state $C_t$ is already known, the protector only needs to sample the utility state $u$ from $B_t$. The sampled state $s$ is comprised of the sampled utility $u$ and the count $C_t$.

Gibbs sampling~\cite{casella1992explaining} is a Markov chain Monte Carlo (MCMC) algorithm for sampling from multivariate probability distributions. Let $X = (x_1, x_2, \ldots, x_n)$ be a general random vector with $n$ components and with finite support described by the multivariate probability density $p(X)$. Gibbs sampling only requires the conditional probabilities $p(x_i|x_{−i})$ to simulate $X$, where $x_{−i} = (x_j)_{j \neq i}$ denotes the subset of all components of $X$ except component $i$. Gibbs sampling is useful when direct sampling from $p(X)$ is difficult. Suppose $k$ samples of $X = (x_1, x_2, \ldots, x_n)$ should be obtained.

Gibbs sampling works in general to produce these samples using only the conditional probabilities $p(x_i|x_{-i})$. It constructs a Markov chain whose steady-state distribution is given by $p(X)$, so that the samples also follow the distribution $p(X)$. The states of this Markov chain are the possible realizations of $X = (x_1, x_2, \ldots, x_n)$, and a specific state $X_i$ is denoted as $Xi = (x_{i1}, x_{i2}, \ldots, x_{in})$ (there are finitely many such states by our assumption). The transition probabilities of this Markov chain, $Pr(X_j|X_i)$, follow from the conditional probabilities $p(x_i|x_{-i})$. Specifically, $Pr(X_j|X_i) = p(x_l|x_{-l})$ when $x_{jv} = x_{iv}$ for all $v$ not equal to $l$, and is equal to zero otherwise, i.e. the state transitions only change one component of the vector--valued sample at a
time. This Markov chain is reversible (meaning $p(X_i)Pr(X_j|X_i) =
P(X_j)Pr(X_i|X_j), \forall i, j$) so $p(X)$ is its steady--state distribution.

\subsubsection{Applying Gibbs sampling in GMOP}
Let $B_t$ be the probability distribution representing the protector's beliefs about the true utilities at the beginning of round $t \geq 1$; $B_1$ represents the protector's prior beliefs when the game starts. Let $B_t(u)$ denote the probability of the vector of utilities $u$ with respect to the distribution $B_t$. Let $B$ be the prior belief distribution and $B′$ be the posterior
belief distribution. Our Bayesian belief update rule to obtain $B′$
from $B$ and the observation is explicitly:

\begin{align*}
B'(s' = (u,C)) &= \eta \Omega(o|s',a) \sum_{s \in S} T(s'|s,a)B(s)\\
&= \eta \Omega(o|s',a)B(s = (u,C-e_a))
\end{align*}

If $a_t$ and $o_t$ represent the actions that the protector and the extractor choose to take at round $t$:

\begin{align*}
B_t(u) &= \eta B_{t-1}(u) \Omega(o_{t-1}|s = (u,C_t),a_{t-1})\\
& = \eta'B_1(u) \prod_{i=1}^{t-1} \Omega(o_i|s = (u,C_{i+1}),a_i)
\end{align*}

It follows that the posterior belief $B_t$ is proportional to the prior
belief $B_1$ multiplied by the observation probabilities over the entire history. Since there are mn possible utilities, it is impossible to store and update $B_t$ when $m$ and $n$ are large, and thus it is impossible to sample directly from $B_t$. Hence, Gibbs sampling is adopted.

Only the conditional probabilities $p(u_i|u_{-i})$, $\forall i$ in $B_t$ are needed:

\begin{align*}
&p(u_i|u_{-i}) = \eta p(u_i, u_{-i}) = \eta B_t(u_i, u_{-i}) =\\
& \eta'B_1(u_i,u_{-i}) \prod_{j=1}^{t-1} \Omega(o_j|s = (u = (u_i, u_{-i}),C_{j+1}),a_j) =\\
& \eta''B_1(u_i) \prod_{j=1}^{t-1} \Omega(o_j|s = (u = (u_i, u_{-i}),C_{j+1}),a_j)
\end{align*}

This quantity is easy to compute where $B_1(u_i)$ is the prior probability that site $i$ has utility $u_i$. Besides the conditional probability, also a valid $u$ with $B_t(u) > 0$ should be found to initialize Gibbs sampling. Finding such a $u$ is easy in our FQR model because any $u$ with $B_1(u) > 0$ satisfies $B_t(u) > 0$.

\bibliographystyle{plain}
\bibliography{contents/report_citations}
\end{document}